\documentclass[aps,pra,reprint,groupedaddress,twocolumn,showpacs]{revtex4-1}

\usepackage[english]{babel}
\usepackage[english]{isodate}
\usepackage{amsmath,amssymb}
\usepackage{graphicx,color}
\usepackage{paralist}
\usepackage{bm}
\usepackage[breaklinks=true,colorlinks=true]{hyperref}

\newcommand{\dchid}{{{}^{\diamond}\hspace{-0.75pt}\chi{}^{\diamond}}}

\begin{document}
\title{Light propagation in local and linear media:\\ Fresnel-Kummer
  wave surfaces with 16 singular points}


\author{Alberto Favaro} \email{a.favaro@imperial.ac.uk}
\affiliation{The Blackett Laboratory, Department of Physics, Imperial
  College London, London SW7\ 2AZ, UK} \author{Friedrich W.\ Hehl}
\affiliation{Institute for Theoretical Physics, University of Cologne,
  50923 Cologne, Germany} \affiliation{Department of Physics and
  Astronomy, University of Missouri, Columbia, MO\ 65211, USA}
\cleanlookdateon\date{\today}

\begin{abstract}
  It is known that the Fresnel wave surfaces of transparent biaxial
  media have 4 singular points, located on two special directions. We
  show that, in more general media, the number of singularities can
  exceed 4. In fact, a highly symmetric linear material is
  proposed whose Fresnel surface exhibits 16 singular points. Because,
  for every linear material, the dispersion equation is quartic, we
  conclude that 16 is the maximum number of isolated singularities. The
  identity of Fresnel and Kummer surfaces, which holds true for media
  with a certain symmetry (zero skewon piece), provides an elegant
  interpretation of the results. We describe a metamaterial
  realization for our linear medium with 16 singular points. It is
  found that an appropriate combination of metal bars, split-ring
  resonators, and magnetized particles can generate the correct
  permittivity, permeability, and magnetoelectric moduli. Lastly, we
  discuss the arrangement of the singularities in terms of Kummer's
  $16_{6}$ configuration of points and planes. An investigation
  parallel to ours, but in linear elasticity, is suggested for future
  research.
\end{abstract}

% Insert suggested PACS numbers in braces on next line

\maketitle

% Body of paper here - Use proper section commands

\section{Introduction}
In the geometrical optics approximation, the Fresnel (wave) surface
describes the propagation of light inside a transparent material. For
biaxial media, such as Aragonite (orthorhombic CaCO$_3$\hspace{0.5pt}) and Topaz (orthorhombic Al$_{2}$SiO$_{4}$(F,OH)$_{2}$\hspace{0.5pt}), the surface consists of two shells that intersect
at four locations \cite{jenkins1957}. Indeed, the property that the
four singular points are aligned on two axes lends the name to this
class of media. The singularities in the Fresnel surface of a biaxial
crystal give rise to the phenomenon of internal conical refraction
\cite{born1980}. More precisely, a narrow light beam (in vacuum)
striking a plate of crystal along the optical axis is refracted, upon
incidence, into a cone and, upon exit, into a hollow cylinder. The
geometry of wave propagation in biaxial media is a classical topic of
research, first investigated by Fresnel \cite{fresnel1868} in 1821. As
an aside, this explains why the wave surface is also termed Fresnel
surface. Conical refraction in biaxial media was predicted as early as
1832 by Hamilton \cite{hamilton1837} on the basis of Fresnel's
work. It was experimentally demonstrated just a year later by Lloyd,
using a crystal of Aragonite \cite{lloyd1833}.

The rise of metamaterial technology \cite{capolino2009} prompts new
questions in the traditional analysis of Fresnel surfaces. By
designing suitable artificial materials, it is possible to control a
wide range of electromagnetic medium parameters. As a matter of fact,
one can tune with remarkable accuracy not only the permittivity, but
also the permeability and the magnetoelectric response. For
comparison, it is worthwhile to recall that the familiar biaxial
crystals have anisotropic permittivity, but are not magnetic or
magnetoelectric. Typically, more complicated media give rise to more
elaborate Fresnel surfaces, whose properties are still largely
unexplored. Our work intends to answer general questions with respect
to the number and arrangement of the singularities in a Fresnel
surface. Most notably, we put forward a highly symmetric medium that
exhibits $16$ singular points, and discuss a metamaterial realization
for it (Sec.\ \ref{sec:noinf} and Sec.\ \ref{sec:meta}).

Two additional remarks are as follows:
\begin{inparaenum}[(i)]
\item Our analysis is restricted to isolated singular points (more correctly termed `ordinary double points' \cite{griffiths1978}). This limitation is important, since, for example, all points on the Fresnel surface of an isotropic medium are singular.
\item The papers \cite{berry2003,berry2005,berry2006} examine conical refraction in bianisotropic media, and are complementary to our work. In particular, they establish that, for media with absorption, the singularities of the Fresnel surface are connected to a wealth of physical effects. 
\end{inparaenum}
%%%%%%

\subsection{Biaxial medium}
\begin{figure}
\includegraphics[width=0.82\columnwidth]{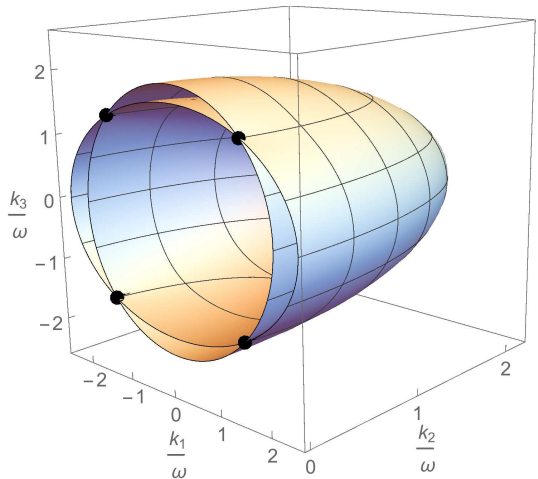}%
\caption{\label{fig:biaxial} (Color online) Cross section of the Fresnel surface for
  the biaxial medium \eqref{eq:biaxial}. The remaining half of the
  surface is obtained via symmetry.}
\end{figure}

A good point of departure is the simple biaxial medium with
electromagnetic response
\begin{equation}
\begin{aligned}
  D^{a}&=\varepsilon^{ab}E_{b}\hspace{0.25pt},\ &\ \,\varepsilon^{ab}&
  =\mbox{diag}(3,4,6),\\
  H_{a}&=\mu^{-1}_{ab}B^{b}\hspace{-0.75pt},\ &\ \,\mu^{-1}_{ab}&
  =\mbox{diag}(1,1,1).\label{eq:biaxial}
\end{aligned}
\end{equation} 
Here, $\varepsilon^{ab}$ and $\mu^{-1}_{ab}$ are the permittivity and
the inverse permeability. Throughout this article, Heaviside-Lorentz units are used, and the vacuum speed of light is normalized to one, whereby $\varepsilon_0=\mu_0=c=1$. Moreover, Latin indices range from $1$ to
$3$, and Einstein's summation convention is employed. The medium
\eqref{eq:biaxial} gives rise to the dispersion equation,
cf. \cite[(D.2.44)]{hehl2003},
\begin{align}
  f(\omega,k_{a})=&-72 \omega^4-3 k_{1}^{4}-4 k_{2}^{4}-6 k_{3}^{4}\nonumber\\
  &+30 \omega^2 k_{1}^{2}+36 \omega^2 k_{2}^{2}+42 \omega^2 k_{3}^{2}\nonumber\\
  &-10 k_{2}^{2} k_{3}^{2}-9 k_{3}^{2}k_{1}^{2}-7 k_{1}^{2}
  k_{2}^{2}\hspace{1pt}=\hspace{1pt}0,\label{eq:biaxialdis}
\end{align}
with $\omega$ being the angular frequency and
$k_{a}=(k_{1},k_{2},k_{3})$ the spatial wave covector. Fig.\
\ref{fig:biaxial} illustrates the Fresnel surface corresponding to
\eqref{eq:biaxialdis}. This surface determines the inverse phase
velocity $k_{a}/\omega$ of an electromagnetic wave traveling in a
given direction. Since the medium is birefringent, there are typically
two possible inverse phase velocities for each direction.

The solutions of the four simultaneous equations 
\begin{align}
  \frac{\partial f(\omega,k_a)}{\partial \omega}&=0, & \frac{\partial
    f(\omega,k_{b})}{\partial k_a}&=0,\label{eq:singular}
\end{align}
are the singular points of the Fresnel surface \cite[\S
79]{landau1960}. In a renowned monograph on Kummer surfaces, Hudson explains this concept as follows \cite[\S 8]{hudson1905}: an isolated singular point is ``characterised geometrically by the fact that the tangent lines at it generate a quadric cone instead of a plane, and algebraically by the absence of terms of the first degree from the equation in point coordinates when the node [the singularity] is taken as origin".

We remark that the solutions of \eqref{eq:singular} can display $\omega=0$, and therefore infinite $k_{a}/\omega$. Hence, representing singular points through the inverse phase velocity
is, at times, impractical. In this article, singular points are
specified via the 4-dimensional wave covector
$q_{\alpha}=(-\omega,k_{a})$, where $\alpha=0,\dots,3$. This
representation does not suffer from the problem just mentioned. It is also worthwhile to observe that multiplying $q_{\alpha}=(-\omega,k_{a})$ by a constant leaves the Fresnel surface unchanged. Thus, if a negative $\omega$ occurs, one is free to multiply the 4-dimensional wave covector by $-1$, and retrieve a positive angular frequency. Because it is possible to take $\omega>0$ systematically, one can introduce a sign convention, and request that the angular frequency must be positive. We do not employ this convention, so that $\omega$ is a generic real number (positive, negative or zero). Retaining the sign flexibility allows us to highlight a pattern in the singular points of Sec.\ \nolinebreak \ref{sec:inf4}, which are interlinked by cyclic permutations of the $4$-dimensional components. 
 
As expected, the locations on the Fresnel surface in Fig.\
\ref{fig:biaxial}, where the two shells meet, are singular
points. Indeed, they correspond to solutions of \eqref{eq:singular}
with $f(\omega,k_a)$ being the function in \eqref{eq:biaxialdis}. The
locations, as represented through $q_{\alpha}$, are
\begin{align*}
(1,+\sqrt{2},0,+\sqrt{2}),& & (1,-\sqrt{2},0,-\sqrt{2}),&\\
(1,+\sqrt{2},0,-\sqrt{2}),& & (1,-\sqrt{2},0,+\sqrt{2}),&
\end{align*}
whereby it is easy to check that the singular points form two optical
axes. In turn, this verifies that the electromagnetic medium
\eqref{eq:biaxial} is biaxial.

The Fresnel surface of the medium \eqref{eq:biaxial} has, in addition
to the 4 real singular points listed above, 12 complex ones, such
as: \begin{equation*} (1,0,\mbox{i}\sqrt{3},\sqrt{6})\,.
\end{equation*}
More generally, the solutions of \eqref{eq:singular} do not need to be
real. The physical interpretation of complex singular points is quite
unclear -- they seem to describe waves that are partly evanescent and whose exponential decay is independent of polarization. A better understanding of complex singular points may be acquired by revisiting the boundary problem of a biaxial medium interfaced with vacuum. This idea is left as the subject of future work.

The discussion so far motivates a question: Can the Fresnel surface of
a linear medium exhibit more than 4 real singular points?

%%%%%%

\section{Sixteen real singular points}
\subsection{Four singular points at infinity}\label{sec:inf4}
A medium is magnetoelectric if an applied electric field induces a
non-zero magnetization and, similarly, an applied magnetic field
induces a non-zero polarization. We consider the medium
\begin{equation}
\begin{aligned}
D^{a}&=\varepsilon^{ab}E_{b}\;+\,\alpha^{a}{}_{b}B^{b},\\
H_{a}&=\mu^{-1}_{ab}B^{b}\hspace{-0.7pt}-\hspace{-0.1pt}\alpha^{b}{}_{a}E_{b},
\end{aligned}\label{eq:magnetoelectric}
\end{equation}
whose magnetoelectric tensor, permittivity, and permeability are
\begin{equation}
\begin{aligned}
  \alpha^{a}{}_{b}&=\textstyle\mbox{diag}\Bigl(\frac{\sqrt{3}}{2},
  -\frac{\sqrt{3}}{2},0\Bigr),\\
  \varepsilon^{ab}&=\textstyle\mbox{diag}\Bigl(1,1,-\frac{1}{2}\Bigr),\\
  \mu^{-1}_{ab}&=\textstyle\mbox{diag}\Bigl(1,1,-\frac{1}{2}\Bigr).
\end{aligned}\label{eq:inf4}
\end{equation}
The dispersion equation specified by \eqref{eq:magnetoelectric} with
\eqref{eq:inf4}, namely
\begin{align}
  f(\omega,k_a)=-\textstyle\frac{1}{2}&\,\Bigl(\;\omega^{4}+k_{1}^{4}+k_{2}^{4}
  +k_{3}^{4}\nonumber\\
  &-\omega^{2} k_{1}^{2}-\omega^{2} k_{2}^{2}-\omega^{2} k_{3}^{2}\nonumber\\
  &-k_{2}^{2} k_{3}^{2}-k_{3}^{2}k_{1}^{2}-k_{1}^{2}
  k_{2}^{2}\hspace{0.75pt}\Bigr)=0,\label{eq:inf4dis}
\end{align}
leads to a Fresnel surface with 16 real singular points. Moreover,
\eqref{eq:inf4dis} is the equation, in homogeneous coordinates, for a
highly symmetric Kummer surface \cite[p.~140]{lord2013}. The general
relationship between Fresnel surfaces and Kummer surfaces is examined
in Sec.\ \ref{sec:kummer}.

By importing $f(\omega,k_a)$ from \eqref{eq:inf4dis} into
\eqref{eq:singular}, one arrives at
\begin{equation}
\begin{aligned}
  \frac{\partial
    f}{\partial\omega}&=\,\omega\hspace{0.5pt}\left(k_{1}^{2}
    +k_{2}^{2}+k_{3}^{2}-2 \omega^{2}\right)=0,\\
  \frac{\partial f}{\partial k_{1}}&=k_{1} \left(\omega^{2}-2
    k_{1}^{2}+k_{2}^{2}+k_{3}^{2}\right)=0,\\
  \frac{\partial f}{\partial k_{2}}&=k_{2} \left(\omega^{2}+k_{1}^{2}
    -2 k_{2}^{2}+k_{3}^{2}\right)=0,\\
  \frac{\partial f}{\partial k_{3}}&=k_{3}
  \left(\omega^{2}+k_{1}^{2}+k_{2}^{2}-2 k_{3}^{2}\right)=0.
\end{aligned}\label{eq:inf4singular}
\end{equation}  
The solutions to \eqref{eq:inf4singular}, that is, the 16 real
singular points of the Fresnel surface for the medium
\eqref{eq:magnetoelectric} with \eqref{eq:inf4} are
\cite{gibbons1993,lord2013}:
\begin{align*}
 \bm{1}\ \ &\ \ (0 , 1 , 1 , 1) &  \bm{9}\ \ &\ \ (0 , 1 , -1 , 1)\\
 \bm{2}\ \ &\ \ (1 , 0 , 1 , 1) & \bm{10}\ \ &\ \ (1 , 0 , 1 , -1)\\
 \bm{3}\ \ &\ \ (1 , 1 , 0 , 1) & \bm{11}\ \ &\ \ (-1 , 1 , 0 , 1)\\
 \bm{4}\ \ &\ \ (1 , 1 , 1 , 0) & \bm{12}\ \ &\ \ (1 , -1 , 1 , 0)\\
 \bm{5}\ \ &\ \ (0 , -1 , 1 , 1) & \bm{13}\ \ &\ \ (0 , 1 , 1 , -1)\\
 \bm{6}\ \ &\ \ (1 , 0 , -1 , 1) & \bm{14}\ \ &\ \ (-1 , 0 , 1 , 1)\\
 \bm{7}\ \ &\ \ (1 , 1 , 0 , -1) & \bm{15}\ \ &\ \ (1 , -1 , 0 , 1)\\
 \bm{8}\ \ &\ \ (-1 , 1 , 1 , 0) & \bm{16}\ \ &\ \ (1 , 1 , -1 , 0)  
\end{align*}

\begin{figure}
  \includegraphics[width=0.82\columnwidth]{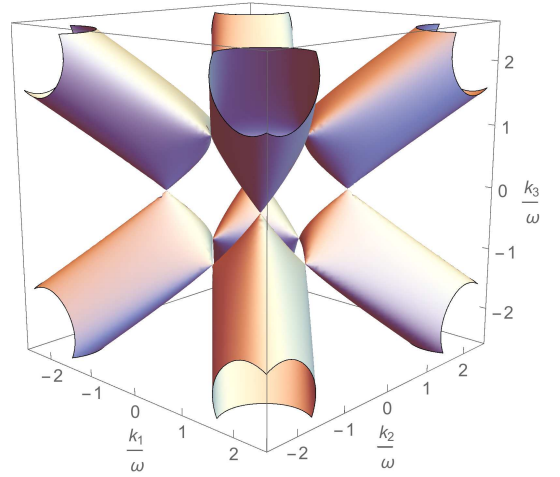}%
 \caption{\label{fig:inf4} (Color online) The Fresnel surface of the magnetoelectric
   medium \eqref{eq:magnetoelectric} with \eqref{eq:inf4} has 12
   singularities in the finite (and 4 at infinity).}
\end{figure}

Fig.\ \ref{fig:inf4} displays the Fresnel surface generated by
\eqref{eq:inf4dis}. A careful inspection of the plot establishes that
it includes only 12 singular points. This is because 4 singular points
are located at infinity. In other words, the electromagnetic waves
described by the covectors $\bm{1},\bm{5},\bm{9}$ and $\bm{13}$ have
$\omega=0$. As a consequence, the inverse phase velocity
$k_{a}/\omega$ for these is unbounded.

It is worthwhile to observe that the 12 singular points included in
Fig.\ \ref{fig:inf4} lie on the edges of a cube, at the
mid-points. The other 4 singularities are retrieved by extending the
body diagonals of the cube to the plane at infinity. As explained in
Sec.\ \ref{sec:sixteensix} below, this arrangement of 16 points has
various remarkable properties.

The question arises: Can the Fresnel surface of a linear medium have
16 real singular points that are all finite?

\subsection{No singular points at infinity}\label{sec:noinf}
We consider the medium specified by \eqref{eq:magnetoelectric}
together with
\begin{align}
  \alpha^{a}{}_{b}&=\textstyle\frac{1}{4}\mbox{diag}
  \bigl(3+\sqrt{3},-3-\sqrt{3},0\bigr),\nonumber\\[0.75pt]\label{eq:noinf}
  \varepsilon^{ab}&=\textstyle\frac{1}{4}\mbox{diag}
  \bigl(-1-\sqrt{3},-1-\sqrt{3},-4+2\sqrt{3}\bigr),\\[0.75pt]
  \mu^{-1}_{ab}&=\textstyle\frac{1}{4}\mbox{diag}
  \bigl(1+\sqrt{3},1+\sqrt{3},4-2\sqrt{3}\bigr),\nonumber
\end{align}
\begin{figure}
\includegraphics[width=0.82\columnwidth]{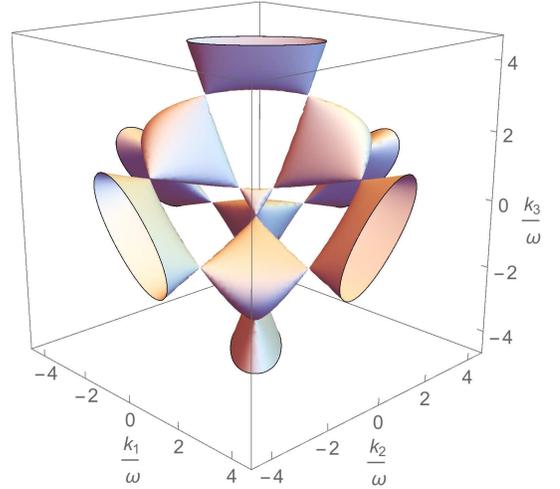}
\caption{\label{fig:noinf} (Color online) The Fresnel surface given by the dispersion equation
  \eqref{eq:noinfdis}. Some readers will note that the plot illustrates a Kummer surface.}
\end{figure}
and formulate the accompanying dispersion equation as
\begin{equation}
f(\omega,k_a)=-\textstyle\frac{1}{2}\bigl(\psi^{2}-6abcd\bigr)=0,
\label{eq:noinfdis}
\end{equation}
where $\psi,a,b,c,d$ are polynomials in the angular frequency and the
spatial wave covector,
\begin{align*}
  \psi(\omega,k_a)&=\textstyle\frac{1}{2}\bigl(11\omega^2
  -k_{1}^2-k_{2}^{2}-k_{3}^{2}\bigr),\\
  a(\omega,k_a)&=\textstyle\frac{1}{2}\bigl(-3\omega-k_1-k_2-k_3\bigr),\\
  b(\omega,k_a)&=\textstyle\frac{1}{2}\bigl(-3\omega-k_1+k_2+k_3\bigr),\\
  c(\omega,k_a)&=\textstyle\frac{1}{2}\bigl(-3\omega+k_1-k_2+k_3\bigr),\\
  d(\omega,k_a)&=\textstyle\frac{1}{2}\bigl(-3\omega+k_1+k_2-k_3\bigr).
\end{align*}

The Fresnel surface of the medium determined by
\eqref{eq:magnetoelectric} and \eqref{eq:noinf} has 16 real singular
points that are all finite. Accordingly, the plot in Fig.\
\ref{fig:noinf} exhibits 16 singularities, which one can view by
applying different rotations. The singular points, as represented via
$q_{\alpha}=(-\omega,k_a)$, are
\begin{align*}
  \bm{1}\ \ &\ \ \textstyle\frac{1}{2}(3,-1,-1,-1) & \bm{9}\ \ &
  \ \ \textstyle\frac{1}{2}(1,1,-3,1)\\
  \bm{2}\ \ &\ \ \textstyle\frac{1}{2}(3,-1 ,1,1) & \bm{10}\ \ &
  \ \ \textstyle\frac{1}{2}(1,1,3,-1)\\
  \bm{3}\ \ &\ \ \textstyle\frac{1}{2}(3,1,-1,1) & \bm{11}\ \ &
  \ \ \textstyle\frac{1}{2}(1,-1,-3,-1)\\
  \bm{4}\ \ &\ \ \textstyle\frac{1}{2}(3,1,1,-1) & \bm{12}\ \ &
  \ \ \textstyle\frac{1}{2}(1,-1,3,1)\\
  \bm{5}\ \ &\ \ \textstyle\frac{1}{2}(1,-3,1,1) & \bm{13}\ \ &
  \ \ \textstyle\frac{1}{2}(1,1,1,-3)\\
  \bm{6}\ \ &\ \ \textstyle\frac{1}{2}(1,1,-1,3) & \bm{14}\ \ &
  \ \ \textstyle\frac{1}{2}(1,-3,-1,-1)\\
  \bm{7}\ \ &\ \ \textstyle\frac{1}{2}(1,3,1, -1) & \bm{15}\ \ &
  \ \ \textstyle\frac{1}{2}(1,-1,1,3)\\
  \bm{8}\ \ &\ \ \textstyle\frac{1}{2}(1,-1,-1,-3) & \bm{16}\ \ & \ \
  \textstyle\frac{1}{2}(1,3,-1,1)
\end{align*}
At these locations, the partial derivatives of the function
$f(\omega,k_a)$ in \eqref{eq:noinfdis} vanish, as is
appropriate. Moreover, because $\textbf{1}$--$\textbf{16}$ have
$\omega\neq 0$, the singular points of the Fresnel surface are indeed
all bounded; to put it differently, the inverse phase velocities of
the electromagnetic waves, which correspond to the 16 singularities, are
finite.

It is interesting to note that, although the surfaces in Fig.\ \ref{fig:biaxial}, Fig.\ \ref{fig:inf4} and Fig.\ \ref{fig:noinf} are dissimilar, the local geometry near each singular point is ultimately the same. Thus, in the geometrical optics limit, there is no fundamental difference between the $16$ singularities above and the $4$ singular points of a biaxial medium. The remark is consistent with algebraic geometry, which treats all isolated singularities on an equal footing.

Now, one may ask: Can the Fresnel surface of a linear medium exhibit 
more than 16 real singular points?\\

\section{Quartic Kummer surfaces}\label{sec:kummer}
The generic local and linear medium is
\begin{equation}
\begin{aligned}
D^{a}&=\varepsilon^{ab}E_{b}\;+\alpha^{a}{}_{b}B^{b},\\
H_{a}&=\mu^{-1}_{ab}B^{b}\hspace{-0.7pt}+\beta_{a}{}^{b}E_{b},
\end{aligned}\label{eq:bianisotropic}
\end{equation}
where
$\{\varepsilon^{ab},\mu^{-1}_{ab},\alpha^{a}{}_{b}\hspace{0.4pt},\beta_{a}{}^{b}\}$
have real components and are
independent of $\{\omega,k_{a}\}$, since the medium is nondispersive.  No further conditions are imposed
on these tensors. As a result,
$\{\varepsilon^{ab},\mu^{-1}_{ab},\alpha^{a}{}_{b}\hspace{0.4pt},\beta_{a}{}^{b}\}$
contain $3\times 3=9$ parameters each, and the medium $4\times 9 = 36$
in total.

The dispersion equation for the generic local and linear response, as
expressed in terms of 3-dimensional quantities, is unwieldy. A more
convenient formula is obtained with the help of 4-dimensional
electrodynamics.  In this representation, the medium
\eqref{eq:bianisotropic} becomes \cite{post1962,hehl2003}:
\begin{equation}\label{eq:medium4d}
\mathcal{H}^{\alpha\beta}={\textstyle\frac{1}{2}}\chi^{\alpha\beta\mu\nu}F_{\mu\nu},
\end{equation}
where the excitation
$\mathcal{H}^{\alpha\beta}\hspace{-1pt}=\hspace{-1pt}-\mathcal{H}^{\beta\alpha}$ and the field
strength $F_{\alpha\beta}\hspace{-1pt}=\hspace{-1pt}-F_{\beta\alpha}$ summarize the fields
$(D^{a},H_{a})$ and $(E_{a},B^{a})$, respectively. Greek indices are
assumed to take the values $\{0,1,2,3\}$. Owing to the symmetries
\begin{equation}
\chi^{\alpha\beta\gamma\delta}=-\chi^{\beta\alpha\gamma\delta}=-\chi^{\alpha\beta\delta\gamma},
\end{equation}
the 4-dimensional medium tensor has 36 independent components, see the
remark above.

One can prove \cite{hehl2003,itin2009,lindell2015} that the dispersion
equation of the generic local and linear medium is quartic in $\omega$ and $k_{a}$. As a matter of fact, in 4 dimensions, it takes the relativistic covariant form
\begin{equation}\label{eq:quartic}
  \tilde{f}(q_{\alpha})=\mathcal{G}^{\kappa\lambda\mu\nu}
  q_{\kappa}q_{\lambda}q_{\mu}q_{\nu}=0,
\end{equation}
with $q_{\alpha}=(-\omega,k_a)$ and
$\tilde{f}(q_{\alpha})=f(\omega,k_a)$. The Tamm-Rubilar tensor,
$\mathcal{G}^{\alpha\beta\gamma\delta}$, may be defined as
\cite{baekler2014}:
\begin{equation}\label{eq:tammrubilar}
  \mathcal{G}^{\alpha\beta\gamma\delta}={\textstyle\frac{1}{3!}}\hspace{1pt}
  \chi^{\kappa(\alpha\beta|\lambda}\hspace{2pt}\dchid_{\!\kappa\mu\lambda\nu}\hspace{1pt}\chi^{\mu|\gamma\delta)\nu}\hspace{0.5pt}.
\end{equation} 
In particular, the indices enclosed by round brackets are
symmetrized. Moreover,
$\dchid_{\!\alpha\beta\gamma\delta}
=\frac{1}{4}\hspace{1pt}
\epsilon_{\alpha\beta\kappa\lambda}\epsilon_{\gamma\delta\mu\nu}\chi^{\kappa\lambda\mu\nu}$,
with $\epsilon_{\alpha\beta\gamma\delta}$ being the Levi-Civita
symbol.

A surface that originates from a quartic homogeneous polynomial
equation in 4 variables cannot have more than 16 isolated singular
points \cite{hudson1905}. The statement refers to the overall number
of singular points---no distinction is made between those that are
real and those that are not. In view of \eqref{eq:quartic}, one
readily concludes that the Fresnel surface of a local and linear
medium never exhibits more than 16 singularities.

By considering index symmetries, one can split the medium tensor into
a principal part, a skewon part, and an axion part \cite{hehl2003}:
\begin{equation}\label{eq:prskax}
  \chi^{\alpha\beta\mu\nu}={^{(1)}}\chi^{\alpha\beta\mu\nu}+{^{(2)}}
  \chi^{\alpha\beta\mu\nu}+{^{(3)}}\chi^{\alpha\beta\mu\nu}. 
\end{equation}
The decomposition is valid in all reference frames, and is irreducible
under the action of $\mbox{GL}(4,\mathbb{R})$. Moreover, the three
elements contain, respectively, $20+15+1=36$ independent
components. When
\begin{equation}\label{eq:noskewon}
\chi^{\alpha\beta\gamma\delta}=\chi^{\gamma\delta\alpha\beta},
\end{equation} 
so that ${^{(2)}}\chi^{\alpha\beta\mu\nu}=0$, the medium is called
skewon-free. Expressing \eqref{eq:noskewon} in 3-dimensional form
yields
\begin{align}
  \varepsilon^{ab}&=\varepsilon^{ba}, & \mu^{-1}_{ab}&=\mu^{-1}_{ba},
  & \beta_{a}{}^{b}&=-\alpha^{b}{}_{a}.
\end{align}
It is then simple to verify that the local linear media with the
constitutive laws \eqref{eq:biaxial}, \eqref{eq:magnetoelectric} with
\eqref{eq:inf4}, and \eqref{eq:magnetoelectric} with \eqref{eq:noinf}
are all skewon-free. Furthermore, as the skewon part of
\eqref{eq:bianisotropic} is arbitrary, this law includes two
independent magnetoelectric tensors, $\alpha^{a}{}_{b}$ and
$\beta_{a}{}^{b}$.

The Fresnel surfaces of skewon-free local and linear media are Kummer
surfaces in the real projective space; conversely, every Kummer
surface is the Fresnel surface of a local and linear medium with
vanishing skewon part \cite{bateman1910,baekler2014}. We deduce that
the Fresnel surface of a skewon-free medium has exactly 16 singular
points, provided these are isolated. As a matter of fact, this
property is known to be valid for the Kummer surfaces
\cite{jessop1903,lord2013}. Notably, the surfaces discussed in earlier
sections had 16 singularities (over the complex numbers), in agreement
with the general theory.

If the electromagnetic response is linear but not local,
$\{\varepsilon^{ab},\mu^{-1}_{ab},\alpha^{a}{}_{b}\hspace{0.4pt},\beta_{a}{}^{b}\}$
can be complex and depend on $\{\omega,k_{a}\}$. The impact of
generalizing the medium parameters to be complex is limited. As the
dispersion equation remains quartic, homogeneous, and polynomial in
nature, the Fresnel surface exhibits no more than 16 singular
points. By contrast, allowing $\{\varepsilon^{ab},\mu^{-1}_{ab},
\alpha^{a}{}_{b} \hspace{0.4pt},\beta_{a}{}^{b}\}$ to depend on
$\{\omega,k_{a}\}$, that is, allowing the material to be dispersive, makes general
questions very difficult.

As an aside, a medium has zero axion part, see \eqref{eq:prskax}, if
in 3 dimensions it satisfies ${\alpha^{a}{}_{a}-
  \beta_{b}{}^{\hspace{0.4pt}b}=0}$ (the indices $a$ and $b$ are summed over). One can easily check that the media
\eqref{eq:biaxial}, \eqref{eq:magnetoelectric} with \eqref{eq:inf4},
and \eqref{eq:magnetoelectric} with \eqref{eq:noinf} are
axion-free. Because the axion part
${^{(3)}}\chi^{\alpha\beta\gamma\delta}$ drops out from the
Tamm-Rubilar tensor \eqref{eq:tammrubilar}, and thus from the
dispersion equation \eqref{eq:quartic}, it is irrelevant for the
propagation of electromagnetic waves \cite{hehl2003}.

\section{Metamaterial realizations}\label{sec:meta}
It appears likely that media with 12 or 16 real and finite singular
points can be realized as metamaterials. More specifically, this
section examines how \eqref{eq:magnetoelectric} with \eqref{eq:inf4},
and \eqref{eq:magnetoelectric} with \eqref{eq:noinf} may be created as
artificial materials.

The permittivity and permeability matrices in \eqref{eq:inf4} and
\eqref{eq:noinf} have the structure $\varepsilon^{ab}=
\mathrm{diag}(\varepsilon_{\perp},\varepsilon_{\perp},
\varepsilon_{\parallel})$ and $\mu^{ab}=\mathrm{diag}(\mu_{\perp},
\mu_{\perp},\mu_{\parallel})$, where the parameters
$\{\varepsilon_{\perp},\varepsilon_{\parallel},\mu_{\perp},\mu_{\parallel}\}$
are allowed to be negative. According to \cite{smith2003},
permittivity and permeability matrices of this type can be achieved
through a suitable arrangement of metal rods and split-ring
resonators. As an aside, the metamaterial will reflect the property
that $\varepsilon^{ab}$ and $\mu^{ab}$ exhibit a preferred direction
(they are uniaxial). Obtaining the numerical values for
$\{\varepsilon_{\perp},\varepsilon_{\parallel},\mu_{\perp},\mu_{\parallel}\}$,
as given in \eqref{eq:inf4} and \eqref{eq:noinf}, requires, almost
certainly, a detailed optimization. More explicitly, the individual
artificial particles must be tuned and their density adjusted. Because one takes
advantage of resonant effects, the desired values of
$\{\varepsilon_{\perp},\varepsilon_{\parallel},\mu_{\perp},\mu_{\parallel}\}$
are expected to be achieved only in a narrow band of frequencies
$\omega$.

To realize magnetoelectric matrices of the type
$\alpha^{a}{}_{b}=\mathrm{diag}(A,-A,0)$, see \eqref{eq:inf4} and
\eqref{eq:noinf}, we suggest two alternative metamaterial designs. The
first design relies on the properties of Chromium Sesquioxide
(Cr$_2$O$_3$), a magnetoelectric crystal \cite{odell1970}. More in
detail, a mixture of perpendicularly arranged Cr$_2$O$_3$ pieces can
be used to generate a matrix $\alpha^{a}{}_{b}$ with the correct
structure. The second design, which is practical at microwave
frequencies, involves a suitable arrangement of magnetostatic wave
resonators \cite{kamenetskii1996,tretyakov1998}. These are ferrite
elements attached to metal wires and magnetized by a static external
magnetic field. As in the case of the permittivity and the
permeability, obtaining a specific numerical value for the
magnetoelectric parameter $A$ is likely to require accurate
optimization. Moreover, one can perform a successful tuning across a
narrow frequency band only.

A further warning is that interactions between the particles associated with different effects (electric, magnetic, and magnetoelectric) may not be negligible. These interactions generally produce unwanted bianisotropic terms in the medium law that are difficult to eliminate. Research on how to address the problem in magnetoelectric metamaterials is ongoing \cite{mirmoosa2014}.

\section{Kummer's $16_{6}$ configuration}\label{sec:sixteensix}
As noted in Sec.\ \ref{sec:inf4}, the singular points of the Fresnel
surface generated by the medium \eqref{eq:magnetoelectric} with
\eqref{eq:inf4} are very symmetric. In fact, it is possible to show
\cite{bruins1959,paerl1975} that the 4-dimensional covectors
$\bm{q}=(-\omega,\bm{k})$ representing these singular locations are
mapped into each other by Dirac $\gamma$-matrices:
\begin{align*}
\underline{\underline{\gamma}}^{0}&=\left(
\begin{array}{cccc}
 1 & 0 & 0 & 0 \\
 0 & 1 & 0 & 0 \\
 0 & 0 & -1 & 0 \\
 0 & 0 & 0 & -1 \\
\end{array}
\right), & 
\underline{\underline{\gamma}}^{1}&=\left(
\begin{array}{cccc}
 0 & 0 & 0 & 1 \\
 0 & 0 & 1 & 0 \\
 0 & -1 & 0 & 0 \\
 -1 & 0 & 0 & 0 \\
\end{array}
\right),\\
\underline{\underline{\gamma}}^{2}&=\left(
\begin{array}{cccc}
 0 & 0 & 0 & -\textrm{i} \\
 0 & 0 & \textrm{i} & 0 \\
 0 & \textrm{i} & 0 & 0 \\
 -\textrm{i} & 0 & 0 & 0 \\
\end{array}
\right), & 
\underline{\underline{\gamma}}^{3}&=\left(
\begin{array}{cccc}
 0 & 0 & 1 & 0 \\
 0 & 0 & 0 & -1 \\
 -1 & 0 & 0 & 0 \\
 0 & 1 & 0 & 0 \\
\end{array}
\right),
\end{align*}
and
$\underline{\underline{\gamma}}^{5}\hspace{-1pt}=\hspace{-1pt}\textrm{i}
\hspace{0.6pt}\underline{\underline{\gamma}}^{0}
\underline{\underline{\gamma}}^{1}\underline{\underline{\gamma}}^{2}
\underline{\underline{\gamma}}^{3}$. Equivalently, one can start from
the 4-dimensional wave covector that describes a singular point, and
retrieve all other singularities by matrix multiplication with
$\underline{\underline{\gamma}}^{0},\dots,
\underline{\underline{\gamma}}^{5}$. For instance, if
$\bm{q}_{1},\dots,\bm{q}_{16}$ are the 16 singular locations listed in
Sec.\ \ref{sec:inf4}, we have \nolinebreak that
\begin{equation}\label{eq:diracmap}
\begin{aligned}
  \bm{q}_{1}=(0,&\,1,1,1), \quad\ & \textrm{i}\hspace{0.6pt}
  \underline{\underline{\gamma}}^{2}\underline{\underline{\gamma}}^{3}\bm{q}_{1}&
  =\bm{q}_{2},\\
  -\underline{\underline{\gamma}}^{0}\bm{q}_{1}&=\bm{q}_{5}, \quad\ &
  -\underline{\underline{\gamma}}^{3}\underline{\underline{\gamma}}^{1}\bm{q}_{1}&
  =\bm{q}_{10},\\
  \underline{\underline{\gamma}}^{1}\bm{q}_{1}&=\bm{q}_{16}, \quad\ &
  -\textrm{i}\hspace{0.6pt}\underline{\underline{\gamma}}^{1}
  \underline{\underline{\gamma}}^{2}\bm{q}_{1}&=\bm{q}_{9},\\
  -\textrm{i}\hspace{0.6pt}\underline{\underline{\gamma}}^{2}\bm{q}_{1}&
  =\bm{q}_{8}, \quad\ & \underline{\underline{\gamma}}^{0}
  \underline{\underline{\gamma}}^{5}\bm{q}_{1}&=\bm{q}_{7},\\
  \underline{\underline{\gamma}}^{3}\bm{q}_{1}&=\bm{q}_{15}, \quad\ &
  -\underline{\underline{\gamma}}^{1}\underline{\underline{\gamma}}^{5}
  \bm{q}_{1}&=\bm{q}_{14},\\
  \underline{\underline{\gamma}}^{0}\underline{\underline{\gamma}}^{1}
  \bm{q}_{1}&=\bm{q}_{4}, \quad\ & \textrm{i}\hspace{0.6pt}
  \underline{\underline{\gamma}}^{2}\underline{\underline{\gamma}}^{5}
  \bm{q}_{1}&=\bm{q}_{6},\\
  \textrm{i}\hspace{0.6pt}\underline{\underline{\gamma}}^{0}
  \underline{\underline{\gamma}}^{2}\bm{q}_{1}&=\bm{q}_{12}, \quad\ &
  -\underline{\underline{\gamma}}^{3}\underline{\underline{\gamma}}^{5}\bm{q}_{1}&
  =\bm{q}_{13},\\
  -\underline{\underline{\gamma}}^{0}\underline{\underline{\gamma}}^{3}
  \bm{q}_{1}&=\bm{q}_{11}, \quad\ &
  \underline{\underline{\gamma}}^{5}\bm{q}_{1}&=\bm{q}_{3}.
\end{aligned}\,
\end{equation}
Here, the extra factors of $-1$ and $\sqrt{-1}$ are immaterial,
because the equations \eqref{eq:inf4singular}, which define the
singular points, are invariant under scaling of $\bm{q}=
(-\omega,\bm{k})$ by any nonzero constant, and the same holds true for
the dispersion equation \eqref{eq:inf4dis}. The striking property
\eqref{eq:diracmap} is a manifestation of the fact that the
singularities give rise to a $16_{6}$ configuration
\cite{lord2013}. More explicitly, the singular points determine $16$
planes, with each plane containing $6$ points; in addition, the planes
meet at the $16$ singular points, with each point lying on $6$ of the
planes.

It is well-known \cite{hudson1905,gibbons1993} that the singularities
of a Kummer surface always identify a $16_{6}$
configuration. Moreover, as explained in Sec.\ \ref{sec:kummer}, the
Fresnel surfaces of skewon-free local and linear media are Kummer
surfaces. We deduce that, for any medium in this wide class, isolated
singular points give rise to a $16_{6}$ configuration.

\section{Conclusion}
We established that rather simple local and linear media can exhibit
16 real singular points. It was found that practical realizations of
these media are within the current technical abilities. To achieve the
required permittivity, permeability, and magnetoelectric moduli,
metamaterial designs were put forward consisting of metal rods,
split-ring resonators, and magnetized inclusions.

On the basis of the general dispersion equation, we discovered that
the Fresnel surface of a local and linear medium cannot exhibit more
than 16 singularities. It was in fact recognized that the Fresnel
surfaces of media with no skewon part have exactly 16 singular points
(assumed isolated). To reach this conclusion, we made use of an
interesting link to the classical projective geometry of Kummer
surfaces.

Finally, it was observed that, for the medium
\eqref{eq:magnetoelectric} with \eqref{eq:inf4}, the singular points
are mapped into each other by the Dirac $\gamma$-matrices. We related
this property to the Kummer $16_6$ configuration and described the
generalization to all local and linear media with vanishing skewon
part.

An idea for future work is to conduct an investigation similar to the
above in linear elasticity. For instance, it appears desirable to
establish how many singularities can the wave surface of a linear
anisotropic elastic medium have. Deriving a general answer to this
question is likely to be more difficult than in electrodynamics. An
upper bound on the number of singularities can however be readily
attained. Because the dispersion equation for anisotropic elastic
media is sextic \cite{itin2013}, the resulting wave surface cannot
display more than 65 singular points; this follows from algebraic
geometry, see \cite{jaffe1997}.

\begin{acknowledgments}
The authors are very grateful to Michael V.\ Berry, Matias F.\ Dahl, Yakov Itin, Ismo V.\ Lindell, Dirk Puetzfeld, Ari H.\ Sihvola, and Sergei A.\ Tretyakov for instructive discussions and comments. A.F.\ would like to thank the Gordon and Betty
Moore Foundation for financial 	\linebreak support.
\end{acknowledgments} 
%\bibliography{singularpts.bib}

\end{document}